\documentclass[runningheads,a4paper]{llncs}

\usepackage{url}
\urldef{\mailsa}\path|{alfred.hofmann, ursula.barth, ingrid.haas, frank.holzwarth,|
\urldef{\mailsb}\path|anna.kramer, leonie.kunz, christine.reiss, nicole.sator,|
\urldef{\mailsc}\path|erika.siebert-cole, peter.strasser, lncs}@springer.com|

\usepackage{color}
\usepackage{graphicx}
\usepackage{mathptmx} 
\usepackage{amsmath, amssymb}
\usepackage{algorithm,algorithmic}
\usepackage{makeidx}  % allows for indexgeneration
\usepackage{graphicx,epsfig,color}
\usepackage{boxedminipage}
\usepackage{multicol} 
\usepackage{enumerate,soul}
\usepackage{diagbox}
\usepackage{float}
\usepackage[table]{xcolor}
\usepackage{latexsym}

\begin{document}

\bibliographystyle{unsrt}	% [1] Firstname Lastname. Article name. ... [in order of reference]

\frontmatter          % for the preliminaries
\pagestyle{headings}  % switches on printing of running heads
\addtocmark{TBD} % additional mark in the TOC
\mainmatter              % start of the contributions
\pagestyle{headings}  % switches on printing of running heads
\addtocmark{Rational Trust Modeling} % additional mark in the TOC
\mainmatter              % start of the contributions
\title{Rational Trust Modeling \\ 
}
\titlerunning{ %Rational Trust Modeling
}  % abbreviated title (for running head)
%\author{Short Paper \vspace{10pt}}
%\authorrunning{ }     % Anonymous Authors

\author{\large Mehrdad Nojoumian %\inst{1} \thanks{Research supported by .} 
%and \inst{2} \thanks{Research supported by .}
}

\authorrunning{ %Nojoumian % \and
}   % abbreviated author list (for running head)

\institute{
Department of Computer \& Electrical Engineering and Computer Science \\ Florida Atlantic University, Boca Raton, Florida, USA \\
\email{mnojoumian@fau.edu} \\
%\and
% \\  \\
%\email{}
}

\maketitle
\begin{abstract}
Trust models are widely used in various computer science disciplines. The primary purpose of a trust model is to continuously measure the trustworthiness of a set of entities based on their behaviors. In this article, the novel notion of \textit{rational trust modeling} is introduced by bridging trust management and game theory. Note that trust models/reputation systems have been used in game theory (e.g., repeated games) for a long time, however, game theory has not been utilized in the process of trust model construction; this is the novelty of our approach. In our proposed setting, the designer of a trust model assumes that the players who intend to utilize the model are rational/selfish, i.e., they decide to become trustworthy or untrustworthy based on the utility that they can gain. In other words, the players are incentivized (or penalized) by the model itself to act properly. The problem of trust management can be then approached by game theoretical analyses and solution concepts such as Nash equilibrium. Although rationality might be built-in in some existing trust models, we intend to formalize the notion of rational trust modeling from the designer's perspective. This approach will result in two fascinating outcomes. First of all, the designer of a trust model can incentivize trustworthiness in the first place by incorporating proper parameters into the trust function, which can be later utilized among selfish players in strategic trust-based interactions (e.g., e-commerce scenarios). Furthermore, using a rational trust model, we can prevent many well-known attacks on trust models. These two prominent properties also help us to predict the behavior of the players in subsequent steps by game theoretical analyses.

\vspace{11pt}
\end{abstract}

%\addtolength{\parskip}{1.6ex}
%\parindent 0pt

\section{Introduction} 
\label{RTM_Introduction}
The primary purpose of a trust model is to continuously measure the trustworthiness of a set of entities (e.g., servers, sellers, agents, nodes, robots, players, etc) based on their behaviors. Indeed, scientists across various disciplines have conducted research on trust over decades and produced fascinating discoveries, however, there is not only a huge gap among findings in these research communities but also these discoveries have not been properly formalized to have a better understanding of the notion of trust, and consequently, practical computational models of trust. We therefore intend to look at the problem of trust modeling from an interdisciplinary perspective that is more realistic and closer to human comprehension of trust.

From a social science perspective, \textit{trust} is the willingness of a person to become vulnerable to the actions of another person irrespective of the ability to control those actions \cite{mayer1995integrative}. However, in the computer science community, \textit{trust} is defined as a personal expectation that a player has with respect to the future behavior of another party, i.e., a personal quantity measured to help the players in their future dyadic encounters. On the other hand, \textit{reputation} is the perception that a player has with respect to another player's intention, i.e., a social quantity computed based on the actions of a given player and observations made by other parties in an electronic community that consists of interacting parties such as people or businesses \cite{mui2002notions}.

From another perspective \cite{castelfranchi1998principles}, \textit{trust} is made up of underlying beliefs and it is a function based on the values of these beliefs. Similarly, \textit{reputation} is a social notion of trust. In our lives, we each maintain a set of reputation values for people we know. Furthermore, when we decide to establish an interaction with a new person, we may ask other people to provide recommendations regarding the new party. Based on the information we gather, we form an opinion about the reputation of the new person. This decentralized method of reputation measurement is called \textit{referral chain}. Trust can be also created based on both local and/or social evidence. In the former case, trust is built through direct observations of a player whereas, in the latter case, it is built through information from other parties. It is worth mentioning that a player can gain or lose her reputation not only because of her cooperation/defection in a specific setting but also based on the ability to produce accurate referrals. 

Generally speaking, the goal of reputation systems is to collect, distribute and aggregate feedback about participants' past behavior. These systems address the development of reputation by recording the behavior of the parties, e.g., in e-commerce, the model of reputation is constructed from a buying agent's positive or negative past experiences with the goal of predicting how satisfied a buying agent will be in future interactions with a selling agent. The ultimate goal is to help the players decide whom to trust and to detect dishonest or compromised parties in a system \cite{resnick2000reputation}. There exist many fascinating applications of trust models and reputation systems in various engineering and computer science disciplines. 

In fact, trust models are widely used in scientific and engineering disciplines such as electronic commerce \cite{DBLP:journals/ci/ZhangCL12,DBLP:conf/trustcom/LiuDFZ12,DBLP:conf/ifiptm/ZhangJZN12,DBLP:conf/pst/GornerZC11,nojoumian2008new,DBLP:journals/dss/JosangIB07}, computer security and rational cryptography \cite{Nojoumian2012,DBLP:conf/gamesec/NojoumianS12,DBLP:conf/pst/NojoumianS12,DBLP:journals/tnsm/FungZB12,NojoumianSG10,NojoumianS10podc}, vehicular ad-hoc networks \cite{li2012reputation,zhang2011survey}, social \& semantic web \cite{gorner2012improving,zhang2011intrank}, multiagent systems \cite{DBLP:journals/taas/WangS10,DBLP:conf/ijcai/WangS07,DBLP:conf/aaai/WangS06}, robotics and autonomous systems \cite{aitken2016assurances,winter2015indian}, game theory and economics \cite{mailath2006repeated,mui2002computational}. To the best of our knowledge, there is no literature on \textit{rational trust modeling}, that is, using game theory during the construction of a trust model. Note that game theoretic models have been used for management and analyses of trust-based systems \cite{feng2014incentive,harish2007game}. 

\subsection{Our Motivation and Contribution}
As our motivation, we intend to provide a new mechanism for trust modeling by which:
\begin{enumerate}
   \item The trust model incentivizes trustworthiness in the first place, i.e., \textit{self-enforcing}.

   \vspace{3pt}
   \item The model is naturally resistant to attacks on trust models, i.e., \textit{resistant}.
\end{enumerate}  

\noindent We therefore introduce the novel notion of \textit{rational trust modeling} by bridging trust management and game theory. We would like to emphasize that trust models have been used in game theory for a long time, for instance, in repeated games to incentivize the players to be cooperative and not to deviate from the game's protocol. However, game theory has not been utilized in the process of trust model construction; in fact, this is the novelty of our proposed approach. 

In our setting, the designer of a trust model assumes that the players who intend to utilize the model are rational/selfish meaning that they cooperate to become trustworthy or defect otherwise based on the utility (to be defined by the trust model) that they can gain, which is a reasonable and standard assumption. In other words, the players are incentivized (or penalized) by the model itself to act properly. The problem of trust modeling can be then approached by strategic games among the players using utility functions and solution concepts such as Nash equilibrium.

Although rationality might be built-in in some existing trust models, we formalize the notion of rational trust modeling from the model designer's perspective. This approach results in two fascinating outcomes. First of all, the designer of a trust model can incentivize trustworthiness in the first place by incorporating proper parameters into the trust function, which can be later utilized among selfish players in strategic trust-based interactions (e.g., e-commerce scenarios between sellers and buyers). Furthermore, using a rational trust model, we can prevent many well-known attacks on trust models, as we describe later. These two prominent properties also help us to predict behavior of the players in subsequent steps by game theoretical analyses.

\subsection{Our Approach in Nutshell}
Suppose there exist two sample trust functions: The first function $f_1(\mathcal{T}^{p-1}_i, \alpha_i)$ receives the previous trust value $\mathcal{T}^{p-1}_i$ and the current action $\alpha_i$ of a seller $S_i$ (i.e., cooperation or defection) as two inputs to compute the updated trust value $\mathcal{T}^{p}_i$ for the next round. However, the second function $f_2(\mathcal{T}^{p-1}_i, \alpha_i, \ell_i)$ has an extra input value known as the seller's lifetime denoted by $\ell_i$. Using the second trust function, a seller with a longer lifetime will be rewarded (or penalized) more (or less) than a seller with a shorter lifetime assuming that the other two inputs (i.e., current trust value and the action) are the same. In this scenario, ``reward" means gaining a higher trust value and becoming more trustworthy, and ``penalty" means otherwise. In other words, if two sellers $S_i$ and $S_j$ both cooperate $\alpha_i=\alpha_j=\mathcal{C}$ and their current trust values are equal $\mathcal{T}^{p-1}_i=\mathcal{T}^{p-1}_j$ but their lifetime parameters are different, for instance, $\ell_i > \ell_j$, the seller with a higher lifetime parameter, gains a higher trust value for the next round, i.e., $\mathcal{T}^{p}_i > \mathcal{T}^{p}_j$. This may help $S_i$ to sell more items and accumulate more revenue because buyers always prefer to buy from trustworthy sellers, i.e., sellers with a higher trust value. 

Now consider a situation in which the sellers can sell defective versions of an item with more revenue or non-defective versions of the same item with less revenue. If we utilize the first sample trust function $f_1$, it might be tempting for a seller to sell defective items because he can gain more utility. Furthermore, the seller can return to the community with a new identity (a.k.a, re-entry attack) after selling defective items and accumulating a large revenue. However, if we use the second sample trust function $f_2$, it's no longer in a seller's best interest to sell defective items because if he returns to the community with a new identity, his lifetime parameter becomes zero and he loses all the credits that he has accumulated overtime. As a result, he loses his future customers and a huge potential revenue, i.e., buyers may prefer a seller with a longer lifetime over a seller who is a newcomer. The second trust function not only incentivizes trustworthiness but also prevents the re-entry attack. 

Note that this is just an example of rational trust modeling for the sake of clarification. The second sample function here utilizes an extra parameter $\ell_i$ in order to incentivize trustworthiness and prevent the re-entry attack. In fact, different parameters can be incorporated into trust functions based on the context (whether it's a scenario in e-commerce or cybersecurity and so on), and consequently, different attacks can be prevented, as discussed in Section \ref{RTM_Discussion}. % using proper analysis.

\section{Preliminaries: Game Theory}
\label{RTM_preliminary}
In this section, preliminary materials regarding game-theoretic concepts are provided for further technical discussions.

A \textit{game} consists of a set of \textit{players}, a set of \textit{actions} and \textit{strategies} (i.e., the method of selecting actions), and finally, a \textit{utility function} that is used by each player to compute how much benefit he can gain by choosing a certain action. In \textit{cooperative games}, players collaborate and split the aggregated utility among themselves, that is, cooperation is incentivized by agreement. However, in \textit{non-cooperative games}, players can not form agreements to coordinate their behavior, that is, any cooperation must be self-enforcing. 

\begin{figure}[h!]
	\centering
		\includegraphics[width=0.95\textwidth]{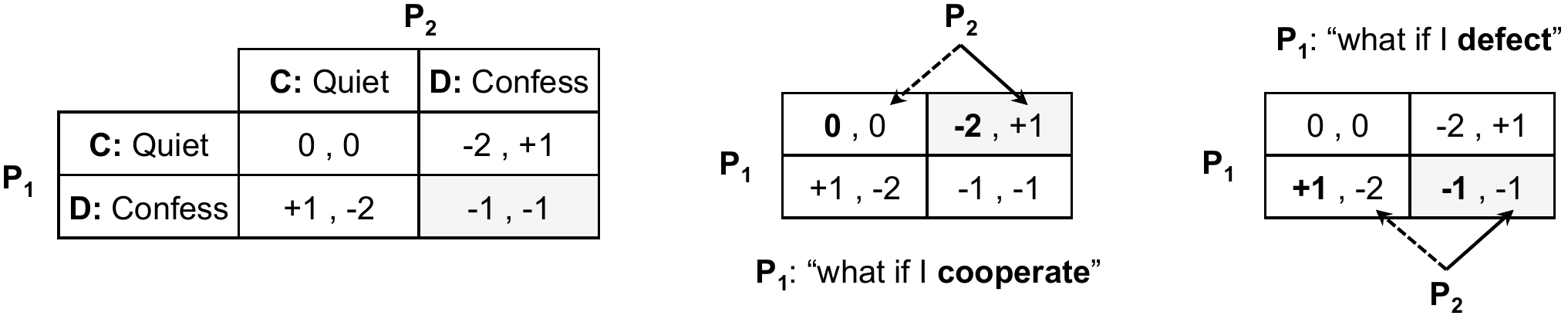}
	\caption{Nash Equilibrium in Prisoner's Dilemma}
	\label{NE_PD}
\end{figure}

The \textit{prisoner's dilemma}, as illustrated in Figure \ref{NE_PD}, is an example of non-cooperative games. In this setting, two possible actions are considered: $\mathcal{C}$: \textit{keep quiet} (or cooperation) and $\mathcal{D}$: \textit{confess} (or defection). In the payoff (utility) matrix, $+1, 0, -1$, and $-2$ denote freedom, jail for one year, jail for two years, and jail for three years respectively. The outcome of this game will be $(\mathcal{D},\mathcal{D})$ because of the \textit{Nash equilibrium} concept, while the ideal outcome is $(\mathcal{C},\mathcal{C})$. To better understand the notion of Nash equilibrium, and consequently, why the game has such an outcome, consider the following two possible scenarios:

\begin{enumerate}
   \item If player $P_1$ selects $\mathcal{C}$ (the first row), then player $P_2$ will select $\mathcal{D}$ (the second column) since $+1>0$.
   
   \vspace{11pt}
   \item If player $P_1$ selects $\mathcal{D}$ (the second row), then player $P_2$ will select $\mathcal{D}$ (the second column) since $-1>-2$.
\end{enumerate}

\noindent In other words, regardless of whether player $P_1$ cooperates or defects, player $P_2$ will always defect. Since the payoff matrix is symmetric, player $P_1$ will also defect regardless of whether $P_2$ cooperates or defects. In fact, since the players are in two different locations and are not able to coordinate their behavior, the final outcome will be $(\mathcal{D},\mathcal{D})$. 

We now briefly review some well-known game-theoretic concepts and definitions \cite{osborne1994course} for our further analysis and discussions.

\begin{definition}
Let $\mathcal{A} \overset{\underset{\mathrm{def}}{}}{=} \mathcal{A}_1 \times \dots \times \mathcal{A}_n$ be an action profile for $n$ players, where $\mathcal{A}_i$ denote the set of possible actions of player $P_i$. A {\em game} $\Gamma=(\mathcal{A}_i,u_i)$ for $1 \leq i \leq n$, consists of $\mathcal{A}_i$ and a utility function $u_i: \mathcal{A} \mapsto \mathbb{R}$ for each player $P_i$. We refer to a vector of actions $\vec{a} = (a_1, \dots, a_n) \in \mathcal{A}$ as an {\em outcome of the game}.
\end{definition}

\begin{definition}
The {\em utility function} $u_i$ illustrates the preferences of player $P_i$ over different outcomes. We say $P_i$ {\em prefers} outcome $\vec{a}$ to $\vec{a}'$ iff $u_i(\vec{a}) > u_i({\vec{a}'})$, and he {\em weakly prefers} outcome $\vec{a}$ to $\vec{a}'$ if $u_i(\vec{a}) \geq u_i({\vec{a}'})$.
\end{definition}

To allow the players to follow randomized strategies (strategy defines how to select actions), we define $\sigma_i$ as a probability distribution over $\mathcal{A}_i$ for a player $P_i$. This means that he samples $a_i \in \mathcal{A}_i$ according to $\sigma_i$. A strategy is said to be a \textit{pure-strategy} if each $\sigma_i$ assigns probability $1$ to a certain action, otherwise, it is said to be a \textit{mixed-strategy}. Let $\vec{\sigma}=(\sigma_1, \dots, \sigma_n)$ be the vector of players' strategies, and let $(\sigma'_i, \vec{\sigma}_{-i}) \overset{\underset{\mathrm{def}}{}}{=} (\sigma_1, \dots, \sigma_{i-1},\sigma'_i, \sigma_{i+1}, \dots, \sigma_n)$, where $P_i$ replaces $\sigma_i$ by $\sigma'_i$ and all the other players' strategies remain unchanged. Therefore, $u_i(\vec{\sigma})$ denote the expected utility of $P_i$ under the strategy vector $\vec{\sigma}$. A player's goal is to maximize $u_i(\vec{\sigma})$. In the following definitions, one can substitute an action $a_i \in \mathcal{A}_i$ with its probability distribution $\sigma_i \in S_i$ or vice versa.

\begin{definition}
A vector of strategies $\vec{\sigma}$ is {\em Nash equilibrium} if, for all $i$ and any $\sigma'_i \neq \sigma_i$, it holds that $u_i(\sigma'_i, \vec{\sigma}_{-i}) \leq u_i(\vec{\sigma})$. This means no one gains any advantage by deviating from the protocol as long as the others follow the protocol.
\end{definition}

\begin{definition}
Let $S_{-i} \overset{\underset{\mathrm{def}}{}}{=} S_1 \times \dots \times S_{i-1} \times S_{i+1} \times \dots \times S_n$. A strategy $\sigma_i \in S_i$ (or an action) is {\em weakly dominated} by a strategy $\sigma'_i \in S_i$ (or another action) with respect to $S_{-i}$ if:
   \begin{enumerate}
      \item For all $\vec{\sigma}_{-i} \in S_{-i}$, it holds that $u_i(\sigma_i, \vec{\sigma}_{-i}) \leq u_i(\sigma'_i, \vec{\sigma}_{-i})$.
      \item There exists a $\vec{\sigma}_{-i} \in S_{-i}$ s.t. $u_i(\sigma_i, \vec{\sigma}_{-i}) < u_i(\sigma'_i, \vec{\sigma}_{-i})$.   
   \end{enumerate}
This means that $P_i$ can never improve its utility by playing $\sigma_i$, and he can sometimes improve it by not playing $\sigma_i$. A strategy $\sigma_i \in S_i$ is {\em strictly dominated} if player $P_i$ can always improve its utility by not playing $\sigma_i$.
\end{definition}

\section{Rational Trust Modeling}
\label{RTM_RTM}
We stress that our goal here is not to design specific trust models or construct certain utility functions. Our main objective is to illustrate the high-level idea of \textit{rational trust modeling} through examples/analyses without loss of generality. %We start with some preliminaries on trust functions.

\subsection{Trust Modeling: Construction and Evaluation}
\label{RTM_Trust_Modeling}
To construct a quantifiable model of trust, a mathematical function or model for trust measurement in a community of $n$ players must be designed. First of all, a basic trust function is defined as follows:

\begin{definition} 
\label{Trust_Fun}
Let $\mathcal{T}^p_i$ denote trust value of player $P_i$ in period $p$ where $-1 \leq \mathcal{T}^p_i \leq +1$ and $\mathcal{T}^0_i=0$ for newcomers. A {\em trust function} is a mapping from $\mathbb{R} \times \mathbb{N}$ to $\mathbb{R}$ which is defined as follows: $(\mathcal{T}^{p-1}_i, \alpha_i) \mapsto \mathcal{T}^p_i$, where $\mathcal{T}^{p-1}_i$ denote the trust value of player $P_i$ in period $p-1$ and $\alpha_i \in \{0,1\}$ denote whether $P_i$ has cooperated, i.e., $\alpha_i = 1$, or defected, i.e., $\alpha_i = 0$, in period $p$.
\end{definition}

\begin{figure}[h!]
	\centering
		\includegraphics[width=0.9\textwidth]{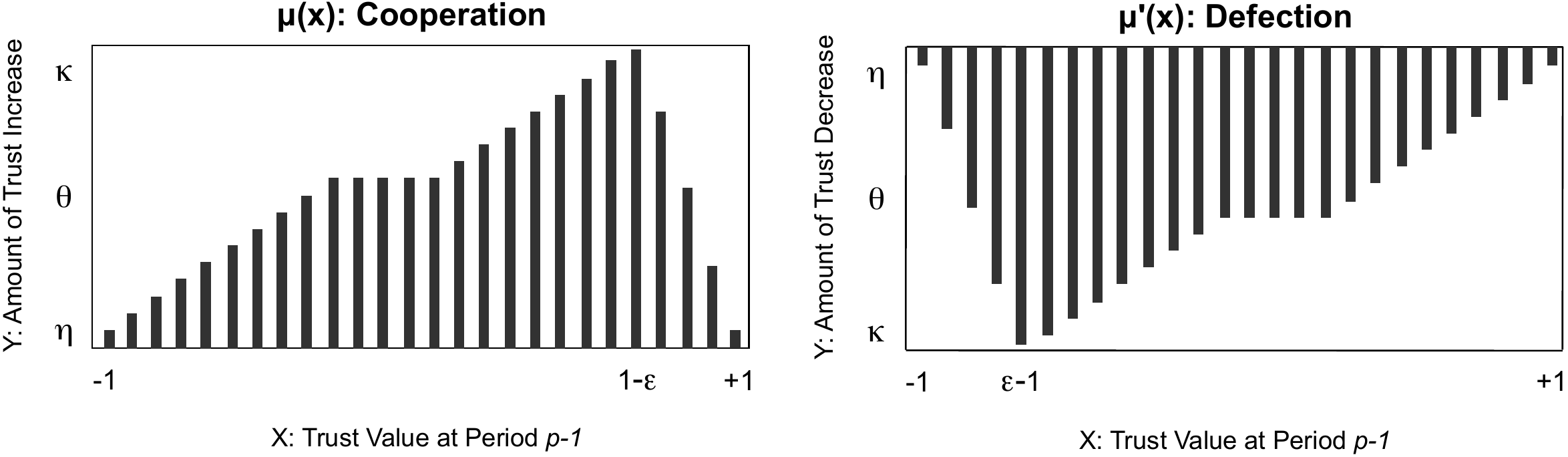}
	\caption{Trust Adjustment by $\mu(x)$ and $\mu'(x)$}
	\label{Trust_Sample_Function}
\end{figure}

As an example, we can refer to the following mathematical model \cite{nojoumian2008new,Nojoumian15wit-ec}. In this model, $\mathcal{T}^p_i = \mathcal{T}^{p-1}_i +\mu(x)$ or $\mathcal{T}^p_i = \mathcal{T}^{p-1}_i -\mu'(x)$ for $\alpha_i=1$ or $\alpha_i=0$ respectively, shown in Figure \ref{Trust_Sample_Function}. Parameters $\eta$, $\theta$ and $\kappa$ are used to reward or penalize players based on their actions (for instance, as defined in \cite{Nojoumian15wit-ec}, $\eta=0.01$, $\theta=0.05$ and $\kappa=0.09$). Note that in $[1-\epsilon,+1]$ and $[-1,\epsilon-1]$, $\mu(x)$ and $\mu'(x)$ both converge to zero, as required by Definition \ref{Trust_Fun}, i.e., $-1 \leq \mathcal{T}^p_i \leq +1$.

After designing a mathematical function, it must be assessed and validated from different perspectives for further improvement. We provide high-level descriptions of some validation procedures to be considered for evaluation of a trust model, that is, \textit{behavioral}, \textit{adversarial} and \textit{operational} methodologies.

\begin{enumerate} %\renewcommand{\labelitemi}{$\bullet$}
   \item \textit{Behavioral}: how the model performs among a sufficient number of players by running a number of standard tests, i.e., executing a sequence of ``cooperation" and ``defection" (or no-participation) for each player. For instance, how fast the model can detect defective behavior by creating a reasonable trust margin between cooperative and non-cooperative parties.
   
   \vspace{3pt}
   \item \textit{Adversarial}: how vulnerable the trust model is to different attacks or any kinds of corruption by a player or a coalition of malicious parties. Seven well-known attacks on trust models are listed below. The first five attacks are known as \textit{single-agent attacks} and the last two are known as \textit{multi-agent} or \textit{coalition attacks} \cite{kerr2013addressing}.
   
   \vspace{3pt} 
%{\small
   \begin{enumerate} %[align=left,label= (2.\arabic*),start=1]
    \item \textbf{Sybil:} forging identities or creating multiple false accounts by one player.
    \vspace{5pt} 
    \item \textbf{Lag:} cooperating for some time to gain a high trust value and then cheat.
    \vspace{5pt} 
    \item \textbf{Re-Entry:} corrupted players return to the scheme using new identities.
    \vspace{5pt}
    \item \textbf{Imbalance:} cooperating on cheap transactions; defecting on expensive ones. 
    \vspace{5pt}
    \item \textbf{Multi-Tactic:} any combination of attacks mentioned above.
    \vspace{5pt}
    \item \textbf{Ballot-Stuffing:} fake transactions among colluders to gain a high trust value.
    \vspace{5pt}
    \item \textbf{Bad-Mouthing:} submitting negative reviews to non-coalition members.   
   \end{enumerate}   
%}   
   \vspace{3pt}
   \item \textit{Operational}: how well the future states of trust can be predicted with a relatively accurate approximation based on possible action(s) of the players (prediction can help us to prevent some well-known attacks), and how well the model can incentivize cooperation in the first place.
\end{enumerate}

\noindent In the next section, we clarify what considerations should be taken into account by the designer in order to construct a proper trust model that resists against various attacks and also encourages trustworthiness in the first place.

\subsection{Rational Trust Modeling Illustration: Seller's Dilemma}
\label{RTM_Sellers_Dilemma}
We now illustrate a dilemma between two sellers by considering two different trust functions. In this setting, each seller has defective and non-defective versions of an item for sale. We consider the following two possible actions:

%\vspace{3pt}
\begin{enumerate}
   \item \textit{Cooperation}: selling the non-defective version of the item for $\$3$ to different buyers.
   
   \vspace{3pt}
   \item \textit{Defection}: selling the defective version of the item for $\$2$ to different buyers.
\end{enumerate}
%\vspace{3pt}

Assuming that the buyers are not aware of the existence of the defective version of the item, they may prefer to buy from the seller who offers the lowest price. This is a pretty natural and standard assumption. As a result, the seller who offers the lowest price has the highest chance to sell the item, and consequently, he can gain more utility. 

\begin{figure}[h!]
	\centering
		\includegraphics[width=1.0\textwidth]{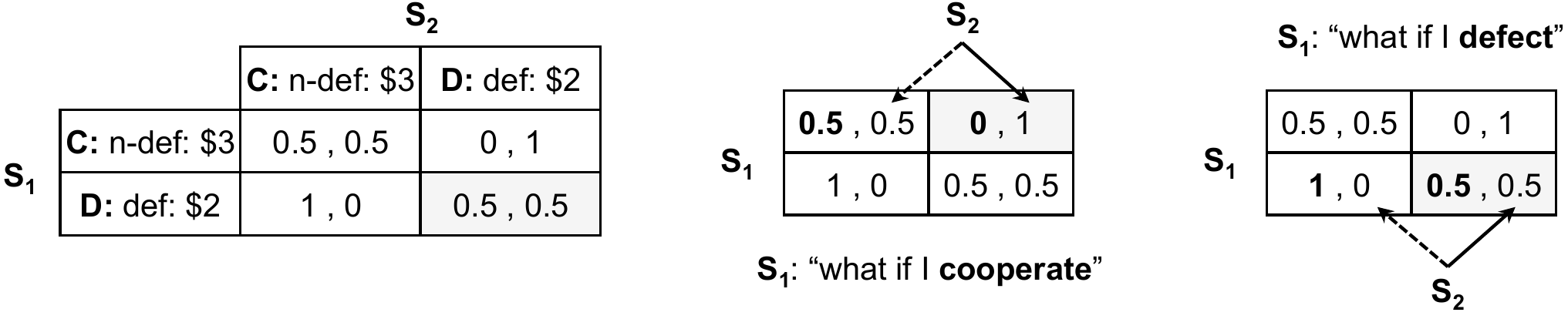}
	\caption{Seller's Dilemma}
	\label{RTM_SD}
\end{figure}

An appropriate payoff function can be designed for this seller's dilemma based on the probability of being selected by a buyer since there is a correlation between the offered price and this probability, Figure \ref{RTM_SD}. In other words, if they both offer the same price ($\$2$ or $\$3$), they have an equal chance of being selected by a buyer, otherwise, the seller who offers a lower price ($\$2$) will be selected by the probability of $1$.

%\vspace{-10pt}
%\vspace{-15pt}

Similar to the prisoner's dilemma, defection is Nash equilibrium meaning that it is in the best interest of each seller to maximize his utility by selling the defective version of the item. For instance, suppose $S_1$ cooperates by selling the non-defective item for $\$3$, $S_2$ will then offer the defective item for $\$2$ to have the highest chance to sell the item, and consequently, he can gain more utility. On the other hand, suppose $S_1$ defects by selling the defective item for $\$2$, $S_2$ will then offer the defective item for $\$2$ to compete with $S_1$. That is, regardless of whether seller $S_1$ cooperates or defects, seller $S_2$ will always defect, and since the payoff matrix is symmetric, defection (selling the defective item) is always Nash equilibrium.

Without loss of generality, we now show how a proper trust model can deal with this dilemma; note that this is just an example for the sake of clarification. We first consider two different trust functions, as we described earlier:

%\vspace{5pt}
\begin{enumerate}
   \item The first function $f_1: (\mathcal{T}^{p-1}_i, \alpha_i) \mapsto \mathcal{T}^p_i$, where $\mathcal{T}^{p-1}_i$ denote the trust value of seller $S_i$ in period $p-1$ and $\alpha_i \in \{0,1\}$ denote whether seller $S_i$ has cooperated or defected in the current period $p$.
   
   \vspace{3pt}
   \item The second function $f_2: (\mathcal{T}^{p-1}_i, \alpha_i, \ell_i) \mapsto \mathcal{T}^p_i$, where $\mathcal{T}^{p-1}_i$ and $\alpha_i \in \{0,1\}$ denote the same notions as of the previous function and $\ell_i \geq 0$ denote the lifetime of seller $S_i$ as a new input in the trust function. This parameter defines how long a seller with a reasonable number of transactions has been in the market.
\end{enumerate}
%\vspace{5pt}

For the sake of simplicity, we didn't consider two different parameters for the lifetime and the number of transactions, however, two separate parameters can be simply incorporated into our trust function and we can still achieve the same game theoretical result. The main reason is because we want to make sure the sellers who have been in the market for a long time but have been inactive or have had a limited number of transactions cannot obtain a high trust value, which is a reasonable assumption.

Now considering the seller's dilemma that we illustrated in Figure \ref{RTM_SD}, the first function $f_1$ is significantly vulnerable to \textit{re-entry} attack. That is, a seller $S_i$ may defect on a sequence of transactions in the middle of his lifetime to gain substantial revenues (utility). He can then return to the market with a new identity as a newcomer. 

However, the lifetime $\ell_i$ is part of the second trust function $f_2$ meaning that a seller $S_i$ with a longer lifetime is more reliable/trustworthy from the buyers' perspective. As a result, he has a higher chance to be selected by the buyers, and consequently, he can gain more utility. This is a very realistic assumption in the e-marketplace. Therefore, it's not in the best interest of a seller to sacrifice his lifetime indicator (and correspondingly his trustworthiness) for a short term utility through defection and then re-entry attack.

It is not hard to show that, by using function $f_2$ rather than function $f_1$, ``defection" is no longer Nash equilibrium in the seller's dilemma, as we illustrate in Section \ref{RTM_Rational_Trust_Modeling}. When we assume the sellers are rational/selfish and they decide based on their utility functions, we can then design a proper trust function similar to $f_2$ to incentivize cooperation in the first place. Furthermore, we can deal with a wide range of attacks, as we mentioned earlier. Finally, at any point, the behavior of a seller can be predicted by estimation of his payoff through trust and utility functions. 

\subsection{Rational Trust Modeling: Design and Analysis} 
\label{RTM_Rational_Trust_Modeling}
In our setting, the utility function $u_i: \mathcal{A} \times \mathcal{T}_i \mapsto \mathbb{R}$, which depends on the seller's action and his trust value. This function computes the utility that each $S_i$ gains or loses by selecting a certain action. If we consider the 2nd trust function $f_2: (\mathcal{T}^{p-1}_i, \alpha_i, \ell_i) \mapsto \mathcal{T}^p_i$, the trust value then depends on the seller's lifetime $\ell_i$ as well. As a result, the lifetime of the seller directly affects the utility that the seller can gain or lose. Now consider the following simple utility function:
\begin{equation}
   u_i = \Omega \times g(\mathcal{T}^{p}_i) \quad \mbox{ where } 0 \leq g(\mathcal{T}^{p}_i) \leq 1, \Omega \mbox{ is a constant}
\end{equation}
As stated earlier, we first define the following parameters, where $-1 \leq \mathcal{T}^{p}_i \leq +1$ and $\alpha_i \in \{0,1\}$ denote whether $S_i$ has cooperated or defected in the previous period:
\begin{equation}
\label{tau}
   \tau_i = \mathcal{T}^{p}_i - \mathcal{T}^{p-1}_i  \quad \mbox{ where } \quad \frac{\left| \tau_i \right|}{\tau_i} = 
   \begin{cases}
   +1 \quad \mbox{ if } \alpha_i=1 \\
   -1 \quad \mbox{ if } \alpha_i=0
   \end{cases} 
\end{equation}

%\vspace{-5pt}
In the following equations, the first function $f_1$ does not depend on the seller's lifetime $\ell_i$, however, the second function $f_2$ has an extra factor that is defined by lifetime $\ell_i$ and constants $\rho$. We can assume that $\rho \ell_i$ is in the same range as of $\mu$ depending on the player's lifetime; that is why $\ell_i$ is multiplied by multiplicative factor $\rho$. Also, it's always positive meaning that no matter if a player cooperates or defects, he will always be rewarded by $\rho \ell_i$. We stress that parameter $\ell_i$ in function $f_2$ is just an examples of how a rational trust function can be designed. The designer can simply consider various parameters (that denote different concepts) as additive or multiplicative factors based on the context in which the trust model is supposed to be utilized. We discuss this issue later in Section \ref{RTM_Discussion} in detail.
\begin{eqnarray}
   f_1 &:& \mathcal{T}^p_i = \mathcal{T}^{p-1}_i + \frac{\left| \tau_i \right|}{\tau_i} \mu \quad \quad \\
   f_2 &:& \mathcal{T}^p_i = \mathcal{T}^{p-1}_i + \frac{\left| \tau_i \right|}{\tau_i} \mu + \rho \ell_i \\ [7pt]  
   -1 &\leq & \mathcal{T}^{p}_i \leq +1, \mbox{ E.g.: } 0 \leq \mu < 0.1 \mbox{ \: is a unified function in $f_1$ and $f_2$} \nonumber %\\ % [10pt]  
\end{eqnarray}

%\vspace{-5pt}
The first function $f_1$ rewards or penalizes the sellers based on their actions and independent of their lifetimes. This makes function $f_1$ vulnerable to different attacks such as the re-entry attack because a malicious seller can always come back to the scheme with a new identity, and then, starts re-building his reputation for another malicious activity. It is possible to make the sign-up procedure costly but it is out of the scope of this paper. 

On the other hand, the second trust function $f_2$ has an extra term that is defined by the seller's lifetime $\ell_i$. This term will be adjusted by $\rho$ as an additional reward or punishment factor in the trust function. In other words, the seller's current lifetime $\ell_i$ in addition to a constant $\pm\beta$ (in the case of cooperation/defection) determine the extra reward/punishment factor. As a result, it is not in the best interest of a seller to reset his lifetime indicator $\ell_i$ to zero because of a short-term utility. This lifetime indicator can increase the seller's trustworthiness, and consequently, his long-term utility overtime.

Let assume our sample utility function is further extended as follows, where $\Omega$ is a constant, for instance, $\Omega$ can be $\$100$:
\begin{equation}
\label{utility}
   u_i = \Omega \bigg( \frac{\mathcal{T}^{p}_i + 1}{2} \bigg) \quad \mbox{ where } 0 \leq \frac{\mathcal{T}^{p}_i + 1}{2} \leq 1, -1 \leq \mathcal{T}^{p}_i \leq +1
\end{equation}

%\vspace{-5pt}
The utility function simply indicates a seller with a higher trust value (which depends on his lifetime indicator as well) can gain more utility because he has a higher chance to be selected by the buyers. In other words, Eqn. (\ref{utility}) maps the current trust value $\mathcal{T}^p_i$ to a value between zero and one, which can be also interpreted as the probability of being selected by the buyers. For the sake of simplicity, suppose $\mathcal{T}^{p-1}_i$ is canceled out in both $f_1$ and $f_2$ as a common factor. The overall utility $U^{f_1}_i$ is shown below when $f_1$ is used. Note that $u_i$ computes the utility of a seller in the case of cooperation or defection whereas $U_i$ also takes into account the \textit{external utility} or \textit{future loss} that a seller may gain or lose. For instance, more savings through selling the defective version of an item instead of its non-defective version.

%\begin{equation*}
%U_i =
%   \begin{cases}
%      +\mu \quad &\mbox{ using } f_1 \mbox{ when } \alpha_i=1 \\ \\
%      -\mu + 3\mu = +2\mu \quad &\mbox{ using } f_1 \mbox{ when } \alpha_i=0 \mbox{ \textbf{plus} }\\ \\
%      &\mbox{ $3\mu$ for selling a defective item } \\ \\
%      +2\mu \quad &\mbox{ using } f_2 \mbox{ when } \alpha_i=1 \\ \\
%      -2\mu + 3\mu = +\mu \quad &\mbox{ using } f_2 \mbox{ when } \alpha_i=0 \mbox{ \textbf{plus} }\\ \\
%      &\mbox{ $3\mu$ for selling a defective item }
%   \end{cases} 
%\end{equation*}
\vspace{-10pt}
\begin{equation*}
U^{f_1}_i = \Omega \times
   \begin{cases}
      \frac{+\mu + 1}{2} \quad &\mbox{ using } f_1 \mbox{ when } \alpha_i=1 \\ 
      \\
      \frac{-\mu + 1}{2} + \beta \quad &\mbox{ using } f_1 \mbox{ when } \alpha_i=0 \mbox{ plus $\beta$, $\beta$ is the \textit{external utility} } \\
      &\mbox{ that the seller obtains by selling the defective item }
   \end{cases} 
\end{equation*}
\vspace{-5pt}

\noindent As shown in $U^{f_1}_i $, function $f_1$ rewards/penalizes sellers in each period by factor $\pm \frac{\mu}{2}$. Accordingly, we can assume \textit{external utility} $\beta$ that the seller obtains by selling the defective item is slightly more than (as much as $\sigma$) the utility that the seller may lose because of defection; otherwise, the seller wouldn't defect, that is, $\beta=\frac{\mu}{2}+|-\frac{\mu}{2}|+\sigma = \mu + \sigma$ (note that the seller not only loses a potential reward $\frac{\mu}{2}$ but also he is penalized by factor $-\frac{\mu}{2}$ when he defects.) In other words, external utility $\beta$ not only compensates for loss $\frac{\mu}{2}+|-\frac{\mu}{2}|$ but also provides additional gain $\sigma$. %; otherwise, the players don't sell the defective item.

As a result, $\frac{-\mu+1}{2}+\beta = \frac{-\mu+1}{2}+(\mu+\sigma) = \frac{\mu+1}{2}+\sigma$. Therefore, $\: \mathcal{D}$efection is always Nash Equilibrium when $f_1$ is used, as shown in Table \ref{table_DD}. We can assume the seller cheats on $\delta$ rounds until he is labeled as an untrustworthy seller. At this point, he leaves and returns with a new identity with the same initial trust value of newcomers, i.e., re-entry attack. Our analysis remains the same even if cheating is repeated for $\delta$ rounds.

\vspace{-15pt}
\renewcommand{\arraystretch}{1.3}
\begin{table} [H]
    \centering
    \begin{tabular}{| c | c | c |}
		\hline
		\textbf{\backslashbox{$S_1$}{$S_2$}} & $\mathcal{C}$ooperation & $\mathcal{D}$efection \\
    \hline \hline
    $\mathcal{C}$ooperation & $\frac{\mu+1}{2},\frac{\mu+1}{2}$ & $\frac{\mu+1}{2},\frac{\mu+1}{2}+\sigma$ \\
    \hline
    $\mathcal{D}$efection & $\frac{\mu+1}{2}+\sigma,+\frac{\mu+1}{2}$ & %\cellcolor{lightgray!50}
    \cellcolor{lightgray!50} $\frac{\mu+1}{2}+\sigma,\frac{\mu+1}{2}+\sigma$ \\
    \hline
    \end{tabular}
    \vspace{7pt}
		\caption{Seller's Dilemma: $\: \mathcal{D}$efection is always Nash Equilibrium using $f_1$.}
		\label{table_DD}
\end{table}

Similarly, function $f_2$ rewards/penalizes sellers through $U^{f_2}_i$ in each period by factor $\pm \frac{\mu}{2}$. Furthermore, this function also has a positive reward (or forgiveness) factor $\frac{\rho \ell_i}{2}$ for cooperative (or non-cooperative) sellers, which is defined by their lifetime factors. Likewise, we can assume \textit{external utility} $\beta$ that the seller obtains by selling the defective item is slightly more than the utility that the seller may lose by defection ($\beta=\mu + \sigma$). The overall utility $U^{f_2}_i$ will be as follows when the $f_2$ is used:
\begin{equation*}
U^{f_2}_i = \Omega \times
   \begin{cases}
      \frac{(+\mu + \rho \ell_i) + 1}{2} \quad &\mbox{ using } f_1 \mbox{ when } \alpha_i=1 \\
      \vspace{5pt} \\
      \frac{(-\mu + \rho \ell_i) + 1}{2} + \beta - \gamma \quad &\mbox{ using } f_1 \mbox{ when } \alpha_i=0 \mbox{ plus $\beta$ as before, } \\
      &\mbox{ where $\gamma$ is the \textit{future loss} due to the impact of $\ell_i$ } %\\
%      &\mbox{ becomes zero after the re-entry attack } \\
   \end{cases} 
\end{equation*}
%\vspace{10pt}

Without loss of generality, suppose the seller defects, leaves and then comes back with a new identity. As a result the lifetime index $\ell_i$ becomes zero. Let assume this index is increased by the following arithmetic progression to reach to where it was: $0 \:,\: \frac{1}{5} \ell_i \:,\: \frac{2}{5} \ell_i \:,\: \frac{3}{5} \ell_i \:,\: \frac{4}{5} \ell_i \:,\: \ell_i$.  In reality, it takes a while for a seller to accumulate this credit based on our definition, i.e., \textit{years of existence} and \textit{number of transactions}. Therefore,
\begin{eqnarray*}
   \gamma &\approx & \frac{\rho}{2} \big( (\ell_i - 0) + (\ell_i - \frac{1}{5} \ell_i) + (\ell_i - \frac{2}{5} \ell_i) + (\ell_i - \frac{3}{5} \ell_i) + (\ell_i - \frac{4}{5} \ell_i) + (\ell_i -  \ell_i) \big) \\
   &=& \frac{\rho}{2} \big( \ell_i + \frac{4}{5} \ell_i + \frac{3}{5} \ell_i + \frac{2}{5} \ell_i + \frac{1}{5} \ell_i + 0 \big) = \frac{3}{2} \rho \ell_i
\end{eqnarray*}
 \vspace{-5pt}
 
\noindent E.g., $(\ell_i - \frac{1}{5} \ell_i)$ denote the \textit{lifetime} could be $\ell_i$, or even more, but it's now $\frac{1}{5} \ell_i$ meaning that the seller is losing $\frac{4}{5} \ell_i$, and so on. We now simplify the $U^{f_2}_i$ when $\alpha_i=0$ as follows:
 \vspace{-10pt}
\begin{eqnarray*}
      U^{f_2}_i &:& \frac{(-\mu + \rho \ell_i) + 1}{2} + \beta - \gamma \\ [-10pt]
      &=& \frac{(-\mu + \rho \ell_i) + 1}{2} + \mu + \sigma - \frac{3}{2} \rho \ell_i = \overbrace{\frac{(+\mu + \rho \ell_i) + 1}{2}}^{\Psi} + \sigma - \frac{3}{2} \rho \ell_i
\end{eqnarray*}
 \vspace{-10pt}
 
This is a simple but interesting result that shows, as long as $\frac{3}{2} \rho \ell_i > \sigma$, $\: \mathcal{C}$ooperation is always Nash Equilibrium when $f_2$ is used, Table \ref{table_CC}. In other words, as long as future loss $\gamma$ is greater than the short-term gain through defection, it's not in the best interest of the seller to cheat and commit to the re-entry attack, that is, the seller may gain a small \textit{short-term} utility by cheating, however, he loses a larger \textit{long-term} utility because it takes a while to reach to $\ell_i$ from $0$. The analysis will be the same if the seller cheats on $\delta$ rounds before committing to the re-entry attack as long as the future loss is greater than the short-term gain. In fact, the role of parameter $\ell_i$ is to make the future loss costly.

 \vspace{-15pt}
\renewcommand{\arraystretch}{1.3}
\begin{table} [H]
    \centering
    \begin{tabular}{| c | c | c |}
		\hline
		\textbf{\backslashbox{$S_1$}{$S_2$}} & $\mathcal{C}$ooperation & $\mathcal{D}$efection \\
    \hline \hline
    $\mathcal{C}$ooperation & %\cellcolor{lightgray!50}
   \cellcolor{lightgray!50} $ \Psi \: , \: \Psi$ & $ \Psi \: , \: \Psi + \sigma -\frac{3}{2} \rho \ell_i$ \\
%    \color{blue} $ \Psi \: , \: \Psi$ & $ \Psi \: , \: \Psi + \sigma -\frac{3}{2} \rho \ell_i$ \\
    \hline
    $\mathcal{D}$efection & $ \Psi + \sigma -\frac{3}{2} \rho \ell_i \: , \: \Psi $ & $ \Psi + \sigma -\frac{3}{2} \rho \ell_i \: , \: \Psi + \sigma -\frac{3}{2} \rho \ell_i$ \\
    \hline
    \end{tabular}
    \vspace{7pt}
		\caption{Seller's Dilemma: $\: \mathcal{C}$ooperation is always NE using $f_2$ when $\frac{3}{2} \rho \ell_i > \sigma$.}
		\label{table_CC}
\end{table}

\section{Technical Analysis and Discussion}
\label{RTM_Discussion}
As stated earlier, we would like to emphasize that our intention here was not to design specific trust models, construct utility functions (which is hard in many cases), target a certain set of attacks, or focus on particular assumptions/games/dilemmas. Our main objective was to illustrate the high-level idea of \textit{rational trust modeling} by some examples and analyses without loss of generality. The presented models, functions, dilemma scenarios, attack strategies, assumptions and parameters can be modified as long as the model designers utilize the technical approach and strategy of rational trust modeling.

As we illustrated, by designing a proper trust function and using a game-theoretical analysis, not only \textit{trustworthiness can be incentivized} but also well-known \textit{attacks on trust functions can be prevented}, such as re-entry attack in our example. Furthermore, \textit{behavior of the players can be predicted} by estimating the utility that each player may gain. % assuming that he will choose a certain action in a specific period.
In this section, we further discuss on these issues while focusing on other types of attacks against trust models. As shown in Table \ref{table_Para}, all single-agent attacks can be simply prevented if the designer of the model incorporates one or more extra parameters (in addition to the previous trust value $\mathcal{T}^{p-1}_i$ and the current action $\alpha_i$) into the function.

\renewcommand{\arraystretch}{1.5}
\begin{table} [h!]
    \centering
    \begin{tabular}{| c | c | c |}
		\hline
		\textbf{Attacks} & \textbf{Parameter} & \textbf{Description} \\
    \hline \hline
    \textbf{Sybil} & Total number of & Prevent the players to create \\
    & Past Transactions & multiple false accounts \\
    \hline
    \textbf{Lag} & High & Prevent the players to cheat  \\
     & Expectancy & after gaining a high trust value \\
    \hline
    \textbf{Re-Entry} & Lifetime & Prevent the players to return \\
     & of the Player & with a new identity \\
    \hline
    \textbf{Imbalance} & Transaction & Prevent the players to cheat \\
    & Cost & on expensive transactions \\
    \hline
    \textbf{Multi Tactic} & Combination of & Prevent the players to defect \\
   & Parameters & in various circumstances \\
    \hline
    \end{tabular}
    \vspace{7pt}
		\caption{Sample Parameters: To deal with single-agent attacks.} % during rational trust modeling.}
		\label{table_Para}
\end{table}

For instance, to deal with the Sybil attack, we can consider a parameter that only reflects the total number of past transactions. In that case, it's not in the best interest of a player to create multiple accounts and divides his total number of transactions among different identities. For imbalance attack, we can consider a parameter for transaction cost, i.e., if the player defects on an expensive transaction, his trust value declines with much faster ratio. For other attacks and their corresponding parameters, see Table \ref{table_Para}.

It is worth mentioning that when the trust value reaches to the saturated region, e.g., very close to $+1$, a player may not have any interest to accumulate more trust credits. However, in this situation, the \textit{high expectancy} parameter (as shown in Table \ref{table_Para}) can be simply utilized in the trust function to warn the fully trusted players that they can sustain this credibility as long as they remain reliable, and if they commit to defections, they will be negatively and significantly (more than others) affected due to high expectancy.

Similarly, we can consider more complicated parameters to incentivize the players not to collude, and consequently, deal with multi-agent/coalition attacks. It is also worth mentioning that consideration should be given to the context in which the trust model is supposed to be used. Some of these parameters are context-oriented and the designer of the model should take this fact into account when designing a rational trust function.

\section{Concluding Remarks}
\label{RTM_conclusion}
In this paper, the novel notion of \textit{rational trust modeling} was introduced by bridging trust management and game theory. In our proposed setting, the designer of a trust model assumes that the players who intend to utilize the model are rational/selfish, i.e., they decide to become trustworthy or untrustworthy based on the utility that they can gain. In other words, the players are incentivized (or penalized) by the model itself to act properly. The problem of trust management can be then approached by strategic games among the players using utility functions and solution concepts such as NE. 

Our approach resulted in two fascinating outcomes. First of all, the designer of a trust model can incentivize trustworthiness in the first place by incorporating proper parameters into the trust function. Furthermore, using a rational trust model, we can prevent many well-known attacks on trust models. These prominent properties also help us to predict the behavior of the players in subsequent steps by game theoretical analyses. As our final remark, we would like to emphasize that our rational trust modeling approach can be extended to any mathematical modeling where some sorts of utility and/or rationality are involved.

%\section{Acknowledgment}
%We would like to thank the anonymous reviewers for their constructive feedback and inspiring comments.

%\vspace{-5pt}
%\bibliographystyle{abbrv}
\renewcommand{\baselinestretch}{1.15} 
\bibliography{RTMRef}

%\section*{Appendix}
% ******* General Function *******

\end{document}